\documentclass[conference]{IEEEtran}
\IEEEoverridecommandlockouts

\usepackage{amsmath,amssymb,amsfonts}
\usepackage{algorithmic}
\usepackage{graphicx}
\usepackage{textcomp}

\usepackage{nohyperref}
\usepackage{url}

\usepackage[style=ieee]{biblatex} %
\usepackage{xspace}
\usepackage{xargs}                  
\usepackage[pdftex,dvipsnames]{xcolor} 
\usepackage{cleveref}
\usepackage{siunitx}
\usepackage{enumitem}

\usepackage{booktabs, threeparttable}
\usepackage{multicol}
\usepackage{multirow}

\def\BibTeX{{\rm B\kern-.05em{\sc i\kern-.025em b}\kern-.08em
    T\kern-.1667em\lower.7ex\hbox{E}\kern-.125emX}}
    
\newcommand{\implementation}{NICER\xspace}

\newcommand{\manipulator}{Image Manipulator\xspace}
\newcommand{\manipulators}{Image Manipulators\xspace}

\newcommand{\assessor}{Quality Assessor\xspace}
\newcommand{\assessors}{Quality Assessors\xspace}

\newcommand{\R}{\mathbb{R}}

\newcommand{\timesmat}{{\mkern-1mu\times\mkern-1mu}}	%

\begin{document}

\title{\fontsize{14}{12} \textbf{\implementation: Aesthetic Image Enhancement with Humans in the Loop}}

\author{\IEEEauthorblockN{Michael Fischer}
\IEEEauthorblockA{\textit{University of Würzburg} \\
Würzburg, Germany \\
\small email: m.fischer@informatik.uni-wuerzburg.de}
\and
\IEEEauthorblockN{Konstantin Kobs}
\IEEEauthorblockA{\textit{University of Würzburg} \\
Würzburg, Germany \\
\small email: kobs@informatik.uni-wuerzburg.de}
\and
\IEEEauthorblockN{Andreas Hotho}
\IEEEauthorblockA{\textit{University of Würzburg}\\
Würzburg, Germany \\
\small email: hotho@informatik.uni-wuerzburg.de}
}

\maketitle

\begin{abstract}

Fully- or semi-automatic image enhancement software helps users to increase the visual appeal of photos and does not require in-depth knowledge of manual image editing.
However, fully-automatic approaches usually enhance the image in a black-box manner that does not give the user any control over the optimization process, possibly leading to edited images that do not subjectively appeal to the user.
Semi-automatic methods mostly allow for controlling which pre-defined editing step is taken, which restricts the users in their creativity and ability to make detailed adjustments, such as brightness or contrast.
We argue that incorporating user preferences by guiding an automated enhancement method simplifies image editing and increases the enhancement's focus on the user. %
This work thus proposes the Neural Image Correction \& Enhancement Routine  (\implementation), a neural network based approach to no-reference image enhancement in a fully-, semi-automatic or fully manual process that is interactive and user-centered.
\implementation iteratively adjusts image editing parameters in order to maximize an aesthetic score based on image style and content.
Users can modify these parameters at any time and guide the optimization process towards a desired direction.
This interactive workflow is a novelty in the field of human-computer interaction for image enhancement tasks.
In a user study, we show that \implementation can improve image aesthetics without user interaction and that allowing user interaction leads to diverse enhancement outcomes that are strongly preferred over the unedited image.
We make our code publicly available to facilitate further research in this direction.

\end{abstract}


\section{Introduction}

With the ever-increasing amount of images taken, it is logical that the casual user neither has the knowledge, time, nor patience to manually edit all images towards pleasing versions.
This, combined with the fact that photography can benefit greatly from image enhancement, 
explains the availability of numerous simple-to-use image enhancement applications.
Fully-automatic enhancement software that can be found in most smartphone photo applications is usually intransparent, leaving users with a result that neither was created in an explainable way nor necessarily correlates with their individual perception of aesthetics.
Semi-automatic approaches often let the users select a single, pre-defined image filter that usually combines different properties, such as higher contrast and higher saturation.
This, evidently, takes control from the user.

We argue that it is beneficial for both, fully- and semi-automatic image enhancement methods, to be able to incorporate the user's individual perception of aesthetics \textbf{before}, \textbf{during}, and \textbf{after} the enhancement process.
We hence propose the \textbf{N}eural \textbf{I}mage \textbf{C}orrection \& \textbf{E}nhancement \textbf{R}outine (\implementation), which allows exactly this.
It consists of two neural network based components:
An \manipulator first applies a set of learned image operations (e.g., contrast, brightness) with variable magnitude onto the unedited source image while a subsequent \assessor then assesses the resulting enhancement quality.
\implementation iteratively optimizes the parameters of the enhancement operations to maximize the \assessor's aesthetic score. 
Due to the iterative approach, users can modify the \manipulator's parameters before, during, and after the optimization process, directing the enhancement procedure towards subjectively more appealing local optima.
While other enhancement tools merely provide preview options for the current filter setting, \implementation's semi-automatic mode allows for an interactive back-and-forth between the user and the automatic optimization and hence facilitates human-computer interaction in image enhancement applications.

Although the flexible architecture of our approach makes it possible to exchange each component with a specifically tailored version (from, e.g., training on a user's photo collection), \implementation can enhance images in a no-reference setting, without any previous info about the user's liking. 
In a user study, we show that the visual appeal of \implementation's fully-automatic enhancement results already is superior to the original images. 
We further show that interweaving user interactions and the automatic enhancement process results in highly diverse images that are subjectively perceived superior.
Our main contributions are
\begin{enumerate}%
    \item \implementation, a novel way of incorporating human aesthetic preferences into the image enhancement process,
    \item a user study assessing \implementation's performance, and
    \item a publicly available repository containing our source code and trained models \cite{nicergithub}.
\end{enumerate}

The rest of this paper is organized as follows:
\Cref{sec:related_work} presents related work on human-centered image enhancement.
\Cref{sec:method} introduces the methodology and components of \implementation.
\Cref{sec:experiments} then assesses \implementation's enhancement quality in a user study.
We conclude this contribution in \Cref{sec:conclusion} and outline starting points for future research.

\begin{figure*}[ht!]
    \centering
    \includegraphics[width=\textwidth]{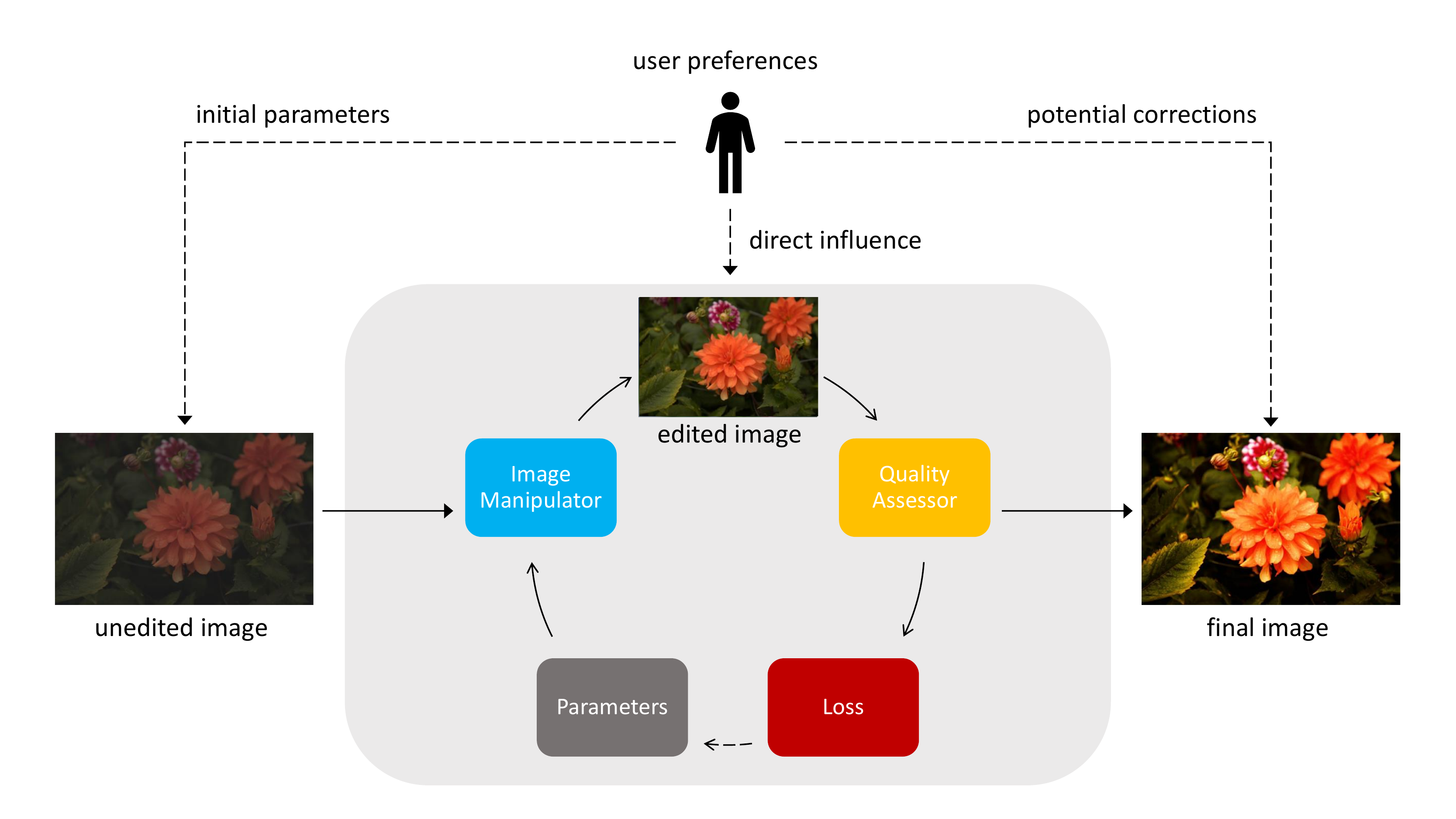}
    \caption{
    \implementation's optimization workflow. If desired, users can interfere with the enhancement routine before, during or after the image optimization. 
    However, user interference is voluntary and not necessary for a successful enhancement.
    The dashed line between loss and parameters implies that the loss does not directly affect the parameters but instead is backpropagated through the architecture.
    Sample image from \cite{fivek}.
    }
    \label{fig:framework-structure}
\end{figure*}

\section{Related Work}
\label{sec:related_work}

The research area of learned perceptual image enhancement has received ample attention in recent works, particularly so after the emergence of neural image assessors \cite{yan2016automatic,nima,fu2018image,kong2016photo}.
However, most approaches do not consider user preferences and enhance the image in a black-box fashion, leaving users with a potentially sub-optimal result that has not been tailored to their personal liking.

The few methods that \textit{do} personalize image enhancement rely on a given photo collection that has already been retouched to the user's liking.
In \cite{kang2010personalization} and \cite{murata2019automatic}, users are directly asked to create this photo collection during the setup phase of their approaches, which is both inconvenient and time-consuming.
Similarly, Hu et al. \cite{hu2018exposure} use Generative Adversarial Networks to learn a latent space from a %
pre-enhanced photo collection and then sample from this space to edit unseen images.
In \cite{kapoor2014collaborative}, styles are grouped into clusters of certain enhancement presets.
New users are then assessed and matched against the enhancement cluster that best suits their preferences.
This approach is most useful for large, cloud-based solutions, where many users are averaged and less suitable for individual, personal image enhancement.

Generally, the mentioned approaches do not explicitly consider the user's individual preferences for the image optimization routine, but rather implicitly use the overall information encoded in the edited image collection.
We argue that such an already edited photo collection is a requirement that might not always be fulfilled (especially for casual users) and further claim that sampling from the style-space might not necessarily yield an edit that is well-suited for a particular image content.
Contrary to the previously mentioned approaches, our method does not rely on a pre-enhanced photo collection that implicitly represents the user's preferences in an abstract style-space.
Although one could train the \assessor to be sensitive to personal preferences by using custom photo collections, this is not necessary for \implementation to work correctly.
Instead, we give the user the freedom to individually guide the optimization by directly interfering with the optimization routine.
The higher amount of (voluntary) interaction can be seen as drawback and benefit at once:
While users might put more effort into getting enhanced images than in fully-automatic enhancement approaches, our method really allows for the individual preferences to be set per image, instead of relying on a globally estimated preferred enhancement style.
Moreover, using a general purpose \assessor and letting users guide the optimization eliminates the need for a pre-enhanced photo collection, making \implementation a ready-to-use, no-reference approach without time-consuming setup.

\section{\implementation}
\label{sec:method}

In this section, we introduce \implementation, our proposed approach for user-centered image enhancement.
The structure of our approach is shown in \Cref{fig:framework-structure} and was motivated by the idea of using a perceptually motivated loss function to increase enhancement appeal \cite{learnedImageEnhancement,doronasca}.
The pipeline consists of the \manipulator, that applies a set of image filters to an unedited source image, and the \assessor, that estimates the aesthetic quality of the \manipulator's outcome.
Using differentiable components allows us to iteratively optimize the \manipulator's filter parameters with respect to the \assessor's score using gradient descent, resulting in a no-reference, automatic image enhancement.
The optimization procedure modifies the filter parameters in each iteration towards the nearest local score optimum and allows user interference in every enhancement step.
If a user changes parameters during optimization, \implementation continues from the new parameter settings towards a different local optimum and thus enables the user to interactively and individually alter the image editing style.

\subsection{\manipulator}

The \manipulator $\psi_{\tiny F}$ is used to apply a set of $n$ image editing filters $\textrm{F} = \{f_1, f_2, ..., f_n\}$ to the image I.
The only requirement of the \manipulator is differentiability with respect to its image filter parameters $\textrm{K} = \{k_1, k_2, ..., k_n\}, \ k_i \in \R$, as these are optimized via gradient descent.
While, in general, any image filter can be used, \implementation implements six common photographic filters: Contrast (Con), Saturation (Sat), Brightness (Bri), Shadows (Sha), Highlights (Hig), Exposure (Exp), and two artistic filters: Local Laplacian Filtering (LLF) \cite{aubry2014fast} and Non-local Dehazing (NLD) \cite{berman2016non}.

We implement the \manipulator as a Context Aggregation Network (CAN), a Fully Convolutional Neural Network with dilated convolutions \cite{yu2015multi} that has been shown to be well-suited for image enhancement tasks \cite{chen2017fast}.
As the CAN is able to approximate a large variety of image filters \cite{chen2017fast}, it is a very flexible and general, yet differentiable enhancement model.

\implementation adapts the CAN24 model by Chen et al. (cf. \Cref{table:can}), as it provides a good trade-off between accuracy and speed~\cite{chen2017fast}.
Between layers one to eight, a leaky rectified linear unit (Leaky ReLU) \cite{xu2015empirical} activation is applied, with a negative slope of $0.2$, while the last layer uses no activation function.
We exclude Batch Normalization, as it showed no significant improvements in approximation accuracy or performance.

Each image filter intensity $k \in \left[-1, 1\right]$ is fed into the network by concatenating it to each pixel of the input image.
During training, we apply one image operation per sample and let the CAN learn the relationship between the input and target output.
At inference time, multiple image filters can be set, as the 
network interpolates correctly and applies the filters simultaneously \cite{chen2017fast}.

In order to learn our proposed image operations, we use the MIT-Adobe FiveK dataset \cite{fivek} with a \num{50}/\num{50} train/test split, resulting in \num{2500} images per set.
We employ the GNU Image Manipulation Program (GIMP) \cite{gimp} to create two manipulated versions (filter intensity $\pm$\num{100}\si{\%}) of each original image for the six photographic filters (Sat, Con, Sha, Hig, Bri, Exp) as ground truth.
To create the ground truth for the filters LLF and NLD, we use the implementations from \cite{aubry2014fast} and \cite{berman2016non}, respectively.
Note that for NLD, we only use positive values (i.e., $+100$\si{\%}), as negative values would haze the image, which is usually undesired in image enhancement.
We then train the \manipulator as in \cite{chen2017fast}.

The trained \manipulator can apply any set of filter intensities onto a source image.
This enables users to initially set or modify filter intensities and provides a way of manually controlling the image editing process, if desired.

\begin{table}[t]
\caption{CAN24 Architecture Overview}

\huge
\begingroup
\setlength{\tabcolsep}{8pt} %
\renewcommand{\arraystretch}{1.2} %
\begin{threeparttable}[t]
\centering
\resizebox{\linewidth}{!}{%
	\begin{tabular}{lccccccccc}
		\toprule[0.25ex]
		\multicolumn{1}{l}{Layer} & \multicolumn{1}{c}{1} & \multicolumn{1}{c}{2} & \multicolumn{1}{c}{3} & \multicolumn{1}{c}{4} & \multicolumn{1}{c}{5} & \multicolumn{1}{c}{6} & \multicolumn{1}{c}{7} & \multicolumn{1}{c}{8} & \multicolumn{1}{c}{9} \\ \hline
		Convolution & $3\timesmat3$ & $3\timesmat3$ & $3\timesmat3$ &  $3\timesmat3$ & $3\timesmat3$ & $3\timesmat3$ & $3\timesmat3 $ & $3\timesmat3$ & $1\timesmat1$ \\
		Dilation & 1 & 2 & 4 & 8 & 16 & 32 & 64 & 1 & 1  \\
		Padding & 1,1 & 2,2 & 4,4 & 8,8 & 16,16 & 32,32 & 64,64 & 1,1 & -  \\
		Receptive Field & $3\timesmat3$ & $7\timesmat7$ & $15\timesmat15$ & $31\timesmat31$ & $63\timesmat63$ & $127\timesmat127$ & $255\timesmat255$ & $257\timesmat257$ & $257\timesmat257$   \\
		Width & 24 & 24 & 24 & 24 & 24 & 24 & 24 & 24 & 3 \\
		\hline
	\end{tabular}}
	\label{table:can}
\end{threeparttable}
\endgroup
\end{table}

\subsection{\assessor}
\label{sec:quality-assessor}

Once the \manipulator $\psi_{\tiny F}$ has edited the image I with the current filter intensity combination K, the \assessor is used as a metric $M$ to rate the manipulation's outcome with a score $S = M(\psi_{\tiny F}(\textrm{K, I}))$, which is then optimized by \implementation.
The \assessor must meet several criteria:
\begin{enumerate}%
    \item Full differentiability with respect to its input.
    \item The \assessor's score prediction must correlate with the human notion of aesthetics.
    This is especially necessary in the automatic enhancement mode, as \implementation will optimize for this score.
    \item $S$ must be deterministic, i.e., same for identical images.
\end{enumerate}

We use a neural network based model called Neural Image Assessment (NIMA) \cite{nima} as \assessor, as it complies with the above desiderata and achieves state-of-the-art performance on aesthetic image assessment.
NIMA fine-tunes a pre-trained Convolutional Neural Network for image classification; in our case VGG16 (conf. D, \cite{Simonyan14c}), as it achieved best cross-dataset performance in \cite{nima}.
The network's output consists of ten nodes that correspond to ten quality score buckets $\{1,2,...,10\}$, where \(10\) is the highest aesthetic rating.
NIMA then feeds the obtained logits through a Softmax function to create a rating distribution, which is the score $S$.

We train NIMA with the Aesthetic Visual Analysis (AVA) dataset \cite{murray2012ava}, whose content ranges from blurry, low-quality snapshots over artistic imagery and advertisements to high-quality photography.
\SI{80}{\%} of the dataset is used for training and the remaining \SI{20}{\%} are equally split into validation and test set.
We follow the training procedure in \cite{nima}.

We find that training solely on the original AVA dataset yields a \assessor that is insensitive to illumination changes, as they are highly under-represented in the dataset.
Therefore, we re-train NIMA's dense layer with \num{3000} images that are manually edited towards bad lightning and whose ground truth scores are decreased to indicate the reduction of aesthetics that comes with poor illumination.

Additionally, we introduce a preprocessing step called Adaptive Brightness Normalization (ABN) to make all initial images have similar brightness.
ABN computes the perceived image brightness $P = \sqrt{0.241 R^2 + 0.691 G^2 + 0.068 B^2}$ using the mean red, green, and blue pixel intensities \cite{hsp}.
With $P \in [0, 255]$, ABN normalizes the brightness to the range $128 \pm 30$.
If the image is too bright ($P > 158$), ABN evens out the original histogram by linearly transforming the pixel values and clipping its left and right side by \SI{5.0}{\%}.
This percentage is reduced if the Structural Similarity \cite{wang2004image} is less than \num{0.8}.
If $P < 98$, ABN brightens up the image by converting it to the HSV color space and increasing the V-value by \num{20}.
As this often introduces unwanted noise, ABN reduces the shift factor if the Peak Signal-to-Noise Ratio \cite{huynh2008scope} between the corrected image and the original version is above \num{30}.
ABN also checks for intended uses of white and black backgrounds, e.g., in product photography.
For this, ABN randomly samples \SI{5}{\%} of the image's pixels and checks if more than \SI{60}{\%} of the sampled pixels are black or white.
If this holds true, the image is not modified to avoid washed out background colors.

\subsection{Optimization Loop}

The overall enhancement process of an image now works as follows:
The image is normalized using ABN and fed through the \manipulator, which applies a set of filters with initial parameters (zero, if not set by user) to the image I.
The image aesthetic of the resulting image is then scored by the \assessor.
We now calculate the gradients of the score w.r.t. the filter parameters K to optimize the parameters via gradient descent towards the nearest local optimum.
This naturally ensures an iterative process a user can interact with before, during, and after the optimization converges.

To optimize the image, \implementation uses a loss function that maximizes image beauty while balancing the ratio between aesthetic gain and induced image change \cite{learnedImageEnhancement}.
We hence formulate the optimization loss as
\begin{equation}
    \mathcal{L}(\textrm{K},\textrm{I}) = \underbrace{\textrm{EMD}(\textrm{\textbf{p}}
	_t, \textrm{\textbf{p}}_d)}_{L_{QA}} \ + \ \underbrace{\gamma \, L_2(\textrm{K})}_{L_{IM}} \,,
\end{equation}
where $L_{QA}$ is the loss that maximizes the \assessor's score and $L_{IM}$ is a weighted regularization term ($\gamma=0.1$) to penalize large parameter changes by the \manipulator, which might not be intended by the user.
More specifically, we define $L_{QA}$ to be the Earth Mover's Distance (EMD) \cite{levina2001earth} between the predicted rating distribution and a desired distribution that corresponds to a highly aesthetic image.
In our implementation, a one-hot encoded target vector for the largest score bucket would force the \assessor to extrapolate towards a ``perfect'' image, which it has never seen during training. 
Therefore, we use a realistic target distribution $\{0.0, 0.0, 0.0, 0.0, 0.0, 0.01, 0.09, 0.15, 0.55, 0.20\}$ that could also be found in the AVA dataset. %
For \assessors that output a single scalar, $L_{QA} = -S$ can be used to maximize the score $S$.
\implementation uses Stochastic Gradient Descent (SGD) with Nesterov Momentum of $0.9$ and a learning rate of $0.05$. %

\subsection{Human in the Loop}

Our approach allows users to intervene with the optimization process at multiple stages:
A user can set initial filter intensities \textbf{before} the optimization loop.
Adding the $L_2$ regularization term to the global enhancement loss ensures that the optimized parameters do not diverge too far from the initial filter intensities set by the user.
Also, by setting the regularization weight $\gamma$, the user can control \implementation's ``diversity'', i.e., the strength with which divergence from the initial filter parameters is penalized. 
A high $\gamma$ could, e.g., be used for fine-tuning an already (subjectively) beautiful image.

A user can modify the filter parameters \textbf{during} the optimization loop, since the gradients are calculated w.r.t. the current parameters.
Setting new parameters may lead to a new local score optimum that the image is optimized towards.
\implementation also provides the option of fixing desired parameter values, which prohibits further intensity updates for a filter and thus ensures that the outcome is to the user's individual liking.

A user can finally modify the parameter settings \textbf{after} the optimization has converged.
This is especially helpful if the optimization yields an image that scores high regarding the \assessor, but tweaks are necessary to increase the image's subjective visual appeal.

\section{Experiments}
\label{sec:experiments}

In this section, we conduct a user study to qualitatively assess the performance of \implementation.
First, we show that the fully-automatic enhancement without any user interaction improves the quality of the original images.
Second, we demonstrate that user interactions can produce different enhancement outcomes that improve the image's subjective appeal.

\subsection{Without User Interaction}

\begin{figure*}[ht!]
    \centering
    \includegraphics[width=\textwidth]{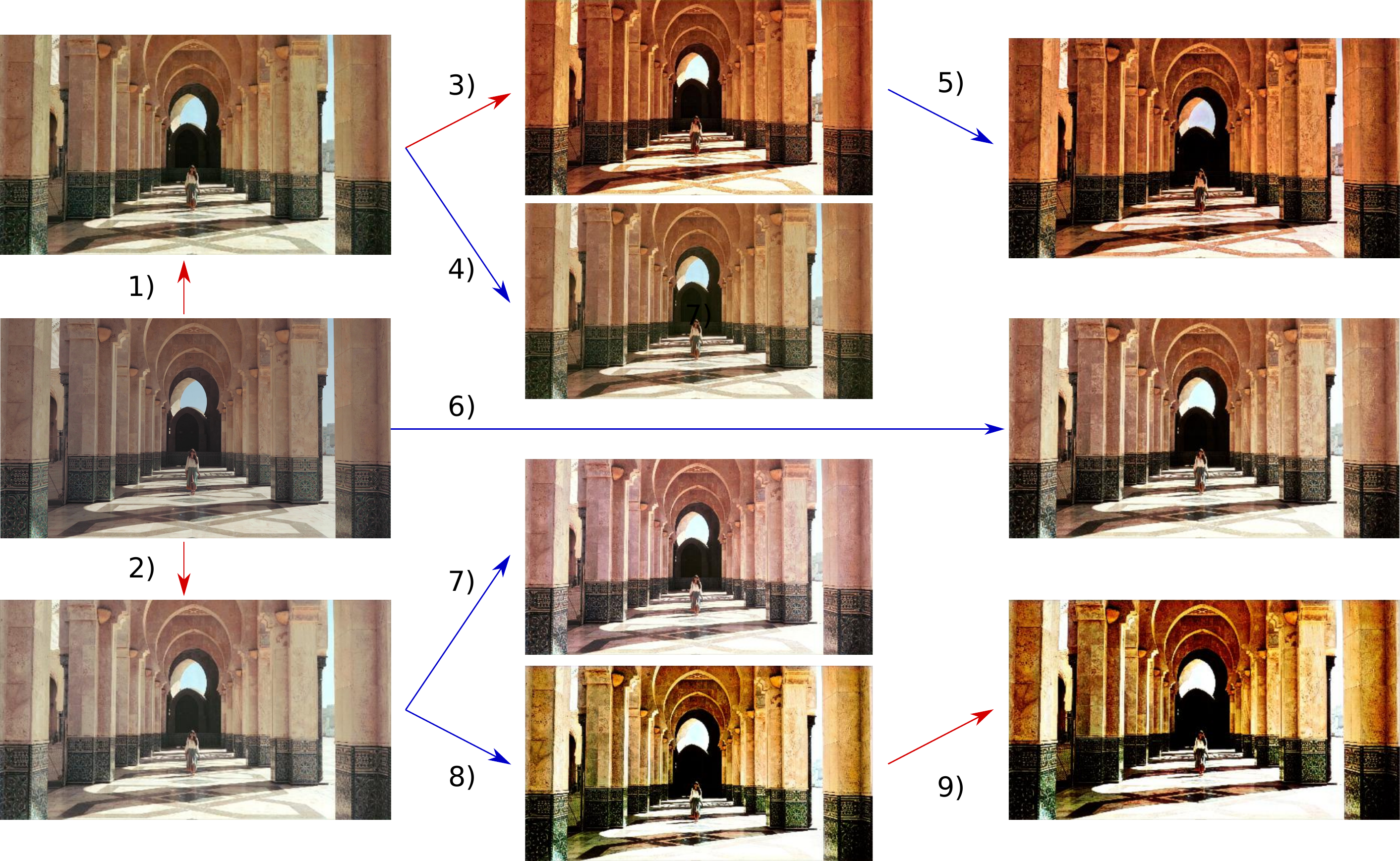}
    \caption{The original image (left column, middle row) with user-defined enhancements (red arrows) and auto-enhancement (blue arrows, default $50$ steps, $\gamma=0.1$). The transformations (given in \%) are: \textbf{1)} Con = Sat = 20, fixed. \textbf{2)} Bri = 8, Sha = -13, Hig = 18, Exp = 24. \textbf{3)} Con = Sat = 20 fixed, NLD = 75 fixed, Exp = 20. \textbf{4)} \& \textbf{6)}: Auto \textbf{5)} Auto, $\gamma=0.5$. \textbf{7)} Auto, stopped by user after $10$ steps. \textbf{8)} NICER, $\gamma=0.005$. \textbf{9)} User post-correction, Hig reduced from 79\% to 40\%. Image from \cite{pexels}.}
    \label{fig:results_interaction}
\end{figure*}

While \implementation is specifically designed to incorporate users into the optimization process, it is also able to automatically enhance images without any interactions.
Showing that fully-automatic enhancement results are perceived as more aesthetic than the original images gives us a ``lower-bound'' that can then be further improved by allowing user interactions.

In the user study, \num{51} subjects sit at a workstation and are instructed to rate \implementation's automatically optimized images (using $\gamma=0.1$), comparing them to the unedited images and versions that are obtained by choosing random filter intensities.
To this end, we use \num{500} randomly sampled images from \cite{fivek} and let each participant rate \num{30} image tuples which results in a total of \num{1530} image ratings.
The subjects are asked to rank the images on a low-to-high scale, with the original reference image centered in the middle of the scale.

\begin{figure}[ht!]
    \centering
    \includegraphics[width=\linewidth]{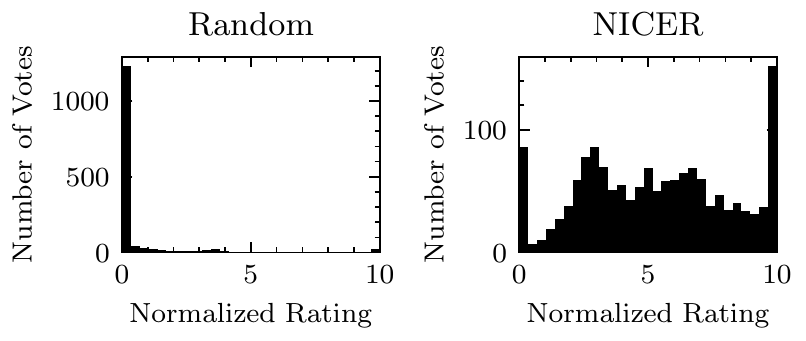}
    \caption{Normalized rating histograms for the random enhancement and NICER's automatic enhancement results. The original image always has a score of \num{5}.}
    \label{fig:ratings}
\end{figure}

The results show that our method is preferred over the random baseline in \SI{93.0}{\%} of all cases.
The subjects prefer our enhancement result over the unedited original image in \SI{53.7}{\%} of all cases.
To quantify the relative ratings, we map the low-to-high scale to the interval $\left[0,10\right]$, where the original image has a score of \num{5} and the other ratings are scaled such that the best or worst rating has a score of \num{0} or \num{10}, respectively.
The normalized rating histograms are shown in \Cref{fig:ratings}.
\implementation's images receive a mean rating of \num{5.3} and a median of \num{5.23}.
A 1-sample Wilcoxon test \cite{rosner2006wilcoxon} shows that the median is significantly different from the original's normalized rating \num{5} in a confidence interval of \SI{1}{\%}.
This suggests that our fully-automatic results on average are perceived more beautiful than the unedited images.
The high variance of the automatic enhancements' rating results ($\sigma^2=2.87$) supports our hypothesis that the perception of image beauty varies greatly across subjects.
To showcase \implementation's full potential, we investigate the effects of directly involving the subjects in the enhancement process.

\subsection{With User Interaction}

We have shown that \implementation can obtain promising results without user interaction.
This section shows how different interactions before, during, and after the optimization loop produce remarkably different enhancement outcomes.
To this end, subjects choose interaction routes from \implementation according to their personal liking, starting from a baseline image.
In \implementation, users have different interaction possibilities: manually change filter settings, fix single filter intensities such that they are not optimized any further, or automatically optimize the image from the current parameter settings for one or multiple steps.
Additionally, they can set the regularization weight $\gamma$.
One set of possible interactions is shown in \Cref{fig:results_interaction}, with not only four different outcomes that involve at least one automatic enhancement step (4, 5, 7, 8), but also the automatically edited version without user interaction (6).
In this experiment, some subjects prefer routes that lead to more saturated looks, while others like high contrasts or slightly tinted images better.
A substantial \SI{97.9}{\%} of the subjects agree that the achieved optimization results are better than the unedited starting image (left column, middle row).
Routes that involve at least one of \implementation's automatic enhancement steps are preferred by \SI{68.1}{\%}.
This shows that combining automatic enhancement with user guidance is a valid approach that yields subjectively more beautiful results.
Fully automatic approaches do not necessarily lead to results that are subjectively aesthetic, which is why  \implementation enables users to intervene with the optimization process at any time and encourages users to bring their individual, preferred style into the enhanced image.

\section{Discussion and Conclusion}
\label{sec:conclusion}

In this paper, we have presented a new method called \implementation to interactively edit and enhance images that allows for the incorporation of user preferences before, during, and after the enhancement process.
The trained models and the source code for \implementation are available online \cite{nicergithub}.

\implementation, being a first implementation of the presented general framework, has certain caveats and weak points that we intend to address in the future:
\implementation runs in reasonable time on a modern machine with a graphics card 
(\SI{1.36}{\second} for the full enhancement of a 1080p image), but might benefit from further optimization when used in hardware-constrained environments like smartphones.
As a first optimization, our enhancement routine rescales images to a width and height of \num{224} pixels during the parameter optimization and only applies the found parameters once to the full-sized image at the end of the optimization cycle.
A further speedup could be achieved by using a more lightweight \assessor.
While the \manipulator could theoretically be replaced by a differentiable image filter library (e.g., Kornia \cite{riba2020kornia}), we explicitly renounced from doing so, as using a neural network makes it possible to not only learn single image filters, but whole editing styles.
This is especially helpful for casual users, who often lack the photographic vocabulary to describe their desired outcome and hence rely more on intuitive terms, such as ``moodiness'', or the indicated settings for a ``sunset'' atmosphere.

We evaluated our results in a user study and found that \implementation's fully-automatic enhancement results usually outperform the unedited images.
In a second experiment, we have shown that \implementation's enhancement results in combination with user interaction were favored by virtually all participants.
In the future, we plan to conduct further in-depth user studies to examine the effects of different \manipulators and \assessors on \implementation's enhancement quality.

Since our approach allows users to intervene with the enhancement process before, during, and after optimization, \implementation offers a first step towards user-centered image editing without reference images.
Overall, we found our method and implementation to be a promising start in this direction.

\printbibliography

\end{document}